\PassOptionsToPackage{unicode}{hyperref}
\PassOptionsToPackage{hyphens}{url}
\PassOptionsToPackage{dvipsnames,svgnames,x11names}{xcolor}
\documentclass[
  12pt]{article}

\usepackage{amsmath,amssymb}
\usepackage{iftex}
\ifPDFTeX
  \usepackage[T1]{fontenc}
  \usepackage[utf8]{inputenc}
  \usepackage{textcomp} 
  \usepackage{multirow}
\else 
  \usepackage{unicode-math}
  \defaultfontfeatures{Scale=MatchLowercase}
  \defaultfontfeatures[\rmfamily]{Ligatures=TeX,Scale=1}
\fi
\usepackage{lmodern}
\usepackage{xcolor}
\usepackage{graphicx}

\makeatletter
\@ifpackageloaded{caption}{}{\usepackage{caption}}

\usepackage[]{natbib}
\usepackage{bookmark}

\usepackage{multirow}

\newcommand{\anon}{1}

\begin{document}

\def\spacingset#1{\renewcommand{\baselinestretch}%
{#1}\small\normalsize} \spacingset{1}


\if1\anon
{
  \title{\bf Bayesian inference for the automultinomial model with an application to landcover data}

  \author{Maria Paula Duenas-Herrera\\
   \footnotesize{mmd6500@psu.edu}\\
   \footnotesize{Department of Statistics - The Pennsylvania State University,}\\
    \footnotesize{Department of Statistics - Universidad Nacional de Colombia}\\
    \footnotesize{and} \\
    Stephen Berg\\
    \footnotesize{sqb6128@psu.edu}\\
    \footnotesize{Department of Statistics - The Pennsylvania State University}\\
    \footnotesize{and}\\
    Murali Haran\\
    \footnotesize{murali.haran@psu.edu}\\
    \footnotesize{Department of Statistics - The Pennsylvania State University}
    }
  \maketitle
} \fi

\if0\anon
{
  \bigskip
  \bigskip
  \bigskip
  \begin{center}
    {\LARGE\bf Title}
\end{center}
  \medskip
} \fi


\bigskip
\begin{abstract}

Multicategory lattice data arise in a wide variety of disciplines such as image analysis, biology, and forestry. We consider modeling such data with the automultinomial model, which can be viewed as a natural extension of the autologistic model to multicategory responses, or equivalently as an extension of the Potts model that incorporates covariate information into a pure-intercept model. The automultinomial model has the advantage of having a unique parameter that controls the spatial correlation. However, the model’s likelihood involves an intractable normalizing function of the model parameters that poses serious computational problems for likelihood-based inference. We address this difficulty by performing Bayesian inference through the Double-Metropolis Hastings algorithm, and implement diagnostics to assess the convergence to the target posterior distribution. 
Through simulation studies and an application to land cover data, we find that the automultinomial model is  flexible across a wide range of spatial correlations  while maintaining a relatively simple specification. For large data sets we find it also has advantages over spatial generalized linear mixed models. To make this model practical for scientists, we provide recommendations for its specification and computational implementation.

\end{abstract}


\section{Introduction}
Lattice data are observations from a random process defined over spatial units arranged in either regular or irregular spatial regions. Examples include county-level rates of Sudden Infant Death Syndrome (SIDS) in North Carolina \citep{Cressie}; remotely sensed satellite imagery of soil composition and topography; and regional species richness in ecology \citep{Zuur2007, Hughes2011}. A key feature of such analyses is the incorporation of spatial dependence among neighboring regions, typically through user-specified neighborhood structures based on geographic proximity, shared boundaries, or other spatial characteristics. An important objective in many lattice data analyses is to explain spatial patterns through covariates measured across the lattice. This is typically accomplished using spatial modeling techniques that account for both observed covariates and the underlying spatial structure. Fitted models can also be used for spatial prediction and interpolation, enabling the estimation of values in unsampled or partially observed regions. While discrete lattice data are well-studied, multicategory lattice data are less explored. This manuscript introduces the automultinomial modeling approach, which explains spatial patterns in multicategory data through covariates while accounting for spatial dependence.\\

Modeling approaches for discrete lattice data generally fall into two categories: conditionally-specified models, also known as automodels \citep{Besag1974}, that directly model the spatial correlation through neighborhood structures, and hierarchical models that rely on latent variables to induce the spatial correlation. In the second group, we have the spatial generalized linear mixed model (SGLMM), a hierarchical model that incorporates spatial dependence through a latent Gaussian Markov Random Field (GMRF). This class of models has a long history, for example \cite{besag1991bayesian} proposed SGLMMs that are widely used for disease mapping. In the context of continuous domain spatial data, \cite{diggle1998model} proposed a general spatial generalized linear mixed model framework for non-Gaussian data; see \cite{haran2011gaussian} for a common framework for both discrete and continuous domain SGLMMs for non-Gaussian data. 
While these models are widely used, they are computationally expensive due to high-dimensional latent variables that leads both to a high-dimensional posterior distribution and slow mixing MCMC algorithms \citep[cf.][]{haran2011gaussian,Hughes2013}, though basis representation approaches \citep{Hughes2013,Haran2025} and fast approximation techniques like INLA \citep{rue2009approximate} can result in efficient inference in many cases. 
Other challenges include spatial confounding \citep{Hodges2012}, which may bias fixed-effect estimates or inflate uncertainty, and some interpretational challenges due to the fact that they describe spatial autocorrelation but rarely explain why the pattern exists \citep{Hughes2017}.\\

For discrete multicategory data, \citet{Imai2005} proposed both conditional and marginal data augmentation strategies for spatial multinomial probit models. Later, \cite{Berrett2012} presented the collapsed and non-collapsed data augmentation strategies for the same model. Although the sampling strategies of \cite{Berrett2012} can lead to gains in computational efficiency, they are not designed to overcome the computational challenges of having a massive amount of data. To address this limitation, \cite{Jin2013} utilizes dimension-reduction techniques to help alleviate the computational burden of a large number of cases. \\ 

We present the automultinomial model as a conditionally specified alternative for modeling discrete multicategory lattice data in the presence of covariate information. \cite{DuenasH2025} studied, within a maximum likelihood framework, the limitations of the pure-intercept specification, known as the Potts model, and showed that it often provides a poor fit for a wide range of spatial configurations. In the presence of strong spatial dependence, the Potts model can exhibit an intrinsic phenomenon known as a phase transition, which restricts its ability to represent realistic spatial arrangements. Motivated by these limitations, we investigate the flexibility of the automultinomial model which is an extension of the Potts model that includes covariates and can thereby 
fit spatial patterns with varying levels of spatial correlation. 
For ease of quantifying uncertainty we pursue a Bayesian approach to fitting the automultinomial model. Because the automultinomial likelihood involves an intractable normalizing function of the parameters  that makes classical likelihood-based inference infeasible, we develop a Double Metropolis–Hastings algorithm and implement diagnostic procedures to assess the quality of the samples produced by the algorithm.\\

The remainder of the paper is organized as follows. Because the automultinomial model is a natural extension of both the autologistic and Potts models, these are reviewed in Sections \ref{sec:autologit} and \ref{sec:Potts} respectively. Section \ref{sec:auxMCMCM} provides an overview of the Double Metropolis–Hastings (DMH) algorithm and a diagnostic for evaluating its quality. Section \ref{sec:simulation} presents the results of two simulation studies. Section \ref{sec:dataanalysis} applies the automultinomial model to a large land cover dataset from Asia originally analyzed in \cite{Berrett2010}. Finally, drawing on insights from both the simulation studies and the real-data application, Section \ref{sec:discussion} discusses the advantages and limitations of the automultinomial model, and provides practical guidance for its use in applied scientific research.

\section{Background} \label{sec:background}

\subsection{The autologistic model}\label{sec:autologit}

The autologistic model is a special case of Besag’s automodels \citep{Besag1974} for binary responses observed on a lattice. Unlike hierarchical spatial models, automodels represent spatial dependence directly through conditional specifications rather than through latent random effects, and define dependence locally via full conditional distributions rather than jointly. The autologistic model has been widely applied in fields such as ecology and epidemiology.\\

Let $\boldsymbol{Y}=(Y_1, \ldots, Y_n)$ denote a binary random field observed on a lattice of size $n$ where $Y_i \in \{0,1\}$ denotes the response at location $i$. Let $\Omega$ represent the set of all possible configurations in the lattice. The traditional autologistic model is defined by the joint distribution 

\begin{equation}\label{eq:autologit}
                    P(\mathbf{Y}=\mathbf{y})=\frac{\exp\{\sum_{i=1}^n \boldsymbol{x}_i^t\boldsymbol{\beta} I(y_i=1)+  \gamma S(\mathbf{y})\}}{\sum_{\mathbf{y} \in \Omega}\exp\{\sum_{i=1}^n \boldsymbol{x}_i^t\boldsymbol{\beta} I(y_i=1) + \gamma S(\mathbf{y})\}},
\end{equation}
where
\begin{equation}\label{eq:sy}
    S(\mathbf{y})=\sum_{i} \sum_{\substack{   i < j \\
       i \sim j}} I(y_i=y_j)
\end{equation}

In Equation (\ref{eq:sy}), $i \sim j$ indicates that areas $i$ and $j$ are neighbors according to a prespecified neighborhood structure. The full conditional distributions for the traditional autologistic model are given by 
$$
\log \left( \frac{\mathbb{P}\left(Y_i=1| Y_j=y_j, j \sim i\right)}{\mathbb{P}\left(Y_i=0| Y_j=y_j, j \sim i\right)} \right)=\mathbf{x}_i^t \boldsymbol{\beta}+ \gamma \sum_{j \sim i} Y_j.
$$

In addition to the traditional specification, \cite{caragea2009} proposed the centered autologistic model, which reparameterizes the spatial term to improve the interpretability of the regression coefficients. This formulation decomposes the model into two components: one that captures the large-scale structure of the process, and another that represents the small-scale spatial dependence in the data. Moreover, because the normalizing constant in \eqref{eq:autologit} is intractable, posterior inference for autologistic models is computationally challenging. In this context, \citet{Hughes2011} provides a comprehensive treatment of Monte Carlo methods for inference in the autologistic model.

\subsection{Potts model}\label{sec:Potts}

Consider a lattice as defined in the previous section. Let $Y_i$ denote the categorical random variable associated with cell $i$ taking values in the set $\{1,2,\ldots, K\}$, and let $\mathbf{Y}$ represent the entire random field. For easier exposition we will henceforth refer to these categories as colors. The probability of observing a particular configuration $\mathbf{y}$ under the Potts model \citep{Potts1952} is
    \begin{equation}\label{eq:PottsModel}
        P(\mathbf{Y}=\mathbf{y})=\frac{\exp\{\sum_{k=2}^{K}\beta_k T_{k} (\mathbf{y})+ \gamma S(\mathbf{y})\}}{\sum_{\mathbf{y} \in \Omega}\exp\{\sum_{k=2}^{K}\beta_k T_{k} (\mathbf{y})+ \gamma S(\mathbf{y})\}},
    \end{equation}
  where
    \begin{equation*}
        T_{k}(\mathbf{y})=\sum_{i}I(y_i=k) \quad     S(\mathbf{y})=\sum_{i} \sum_{\substack{   i < j \\
       i \sim j}} I(y_i=y_j).
    \end{equation*}
We impose the constraint $\beta_1=0$ for identifiability. \\

In Equation (\ref{eq:PottsModel}), $T_{k}(\mathbf{y})$ are sufficient statistics that count the number of cells assigned to each color, while $S(\mathbf{y})$ is a sufficient statistic that counts the number of neighboring cell pairs that share the same color. This statistic therefore quantifies the degree of spatial dependence in the data. The interpretation of the model parameters is as follows: $\beta_k$, with $k=2,\ldots,K$, controls the proportion of cells of color $k$ with respect to the proportion of cells of color $1$. Positive values of $\beta_k$ indicate a higher propensity for color $k$ relative to color $1$, while negative values indicate a lower propensity. On the other hand, the parameter $\gamma$ controls the degree of spatial dependence, with larger values favoring configurations with more agreeing neighboring cells. In order to extend the Potts model to incorporate covariate information, we rewrite Equation (\ref{eq:PottsModel}) as
        \begin{equation}\label{eq:PottsInt}
                P(\mathbf{Y}=\mathbf{y})=\frac{\exp\{\sum_{k=2}^K\sum_{i=1}^{n}\beta_k I(y_i=k)+ \gamma S(\mathbf{y})\}}{\sum_{\mathbf{y} \in \Omega}\exp\{\sum_{k=2}^K\sum_{i=1}^{n}\beta_k I(y_i=k)+ \gamma S(\mathbf{y})\}}.
    \end{equation}
Equation (\ref{eq:PottsInt}) shows that the Potts model can be viewed as an intercept-only model, with one intercept for each category, and a single parameter $\gamma$ that controls the level of spatial correlation in the generated configurations. Because the model does not incorporate any information about the spatial locations of the colors, the goal of modeling in this case is to preserve the overall color proportions and the level of spatial dependence in the arrangement. An extensive analysis of the Potts model for multicategory data on a lattice by \cite{DuenasH2025} studied the issues that arise from phase transitions and stable states in scenarios characterized by high spatial correlation. In particular, simulation studies helped identify situations where this model exhibits lack of fit and becomes inappropriate. The proposed solution is called the Tapered Potts model, which introduces a tapering term that brings flexibility into the model in order to fit a wide range of arrangements, especially when a high level of spatial correlation is present. For cases where the spatial locations of the colors must be preserved, we extend the Potts model by incorporating a set of covariates that provide greater flexibility and ensure preservation of the observed category locations. This extended model is described in Section \ref{sec:automultinomial}.

\section{The automultinomial model} \label{sec:automultinomial}

Consider the arrangement $\mathbf{Y}$ defined in the previous section to be the response of a model built to include a set of $p$ covariates. Our new goal is to model the probability of observing a particular value of $\mathbf{Y}$ with the help of the new available covariates. Let $\boldsymbol{X}$ be a design matrix of size $n \times p$ whose columns correspond to the previously mentioned covariates, and let $\boldsymbol{x}_i$ denote the ith row of $\boldsymbol{X}, i=1,2, \ldots n$. Consider also a parameter matrix $\boldsymbol{\beta}$ of dimension $p \times (K-1)$ (or  $(p+1) \times (K-1)$ if an intercept is included in the model). The column $k-th$ of $\boldsymbol{\beta}$ is denoted by $\boldsymbol{\beta}_k$, and it represents the set of parameters that define the linear predictor associated with the category $k$.
The probability of observing a particular arrangement $\boldsymbol{y}$ is 
\begin{equation}\label{eq:autopmd}
                    P(\mathbf{Y}=\mathbf{y})=\frac{\exp\{\sum_{i=1}^n \sum_{k=2}^{K}\boldsymbol{x}_i^t\boldsymbol{\beta}_k I(y_i=k)+  \gamma S(\mathbf{y})\}}{\sum_{\mathbf{y} \in \Omega}\exp\{\sum_{i=1}^n \sum_{k=2}^{K}\boldsymbol{x}_i^t\boldsymbol{\beta}_k I(y_i=k) + \gamma S(\mathbf{y})\}}.
\end{equation}

In Equation (\ref{eq:autopmd}), the first term in the numerator captures covariate-driven, category-specific effects. The statistic $S(\mathbf{y})$ is defined the same as in Equation (\ref{eq:PottsModel}), and the second term $\gamma S(\mathbf{y})$ is commonly known as an autocovariate. This formulation reduces to the Potts model when the linear predictor contains only intercept terms. \\

The denominator in Equation (\ref{eq:autopmd}) is an intractable normalizing function of $\boldsymbol{\beta}$ and $\gamma$, which will be denoted by $Z(\boldsymbol{\beta}, \gamma)$. The presence of this term in the mass function represents a significant burden in performing maximum-likelihood inference for this model given that $Z(\boldsymbol{\beta}, \gamma)$ cannot be evaluated in a maximization process. One widely known alternative is to maximize an approximation of the likelihood function called the pseudolikelihood function \citep{Besag1975}. However, the maximum pseudolikelihood estimates (MPLE) provide poor  approximations to maximum likelihood estimates (MLE) except when the dependence is weak \citep[cf.][]{Geyer1991}. When performing Bayesian inference via an MCMC algorithm, the acceptance probability at each step of a Metropolis–Hastings sampler requires evaluation of the normalizing function $Z(\boldsymbol{\beta}, \gamma)$, making it impossible to implement. However, several methods have been proposed to enable Bayesian inference in such settings. Section \ref{sec:auxMCMCM} introduces a class of methods known as auxiliary variable MCMC algorithms, including the Double Metropolis–Hastings (DMH) algorithm, which is adopted for inference in this work. Section \ref{sec:ACD} presents a brief description of the Approximate Curvature Diagnostic (ACD) used to assess the adequacy of the DMH sampler. Sampling from the automultinomial model is carried out using the Gibbs sampler implemented in the \textit{automultinomial} package \citep{Berg2019_2}.

\subsection{Double Metropolis-Hastings algorithm} \label{sec:auxMCMCM}
 
The Double Metropolis–Hastings (DMH) algorithm, originally proposed by \cite{Liang2010}, belongs to a class of methods known as auxiliary-variable MCMC \citep{Moller2006}. These methods introduce an auxiliary variable $\boldsymbol{z}$ into a Metropolis-Hastings algorithm for the posterior of the parameters of interest. With an appropriate choice of proposal distribution, the intractable normalizing constants cancel from the Metropolis–Hastings acceptance ratio. We are interested in sampling from the posterior distribution
\begin{equation*}
    \pi(\boldsymbol{\theta}|\boldsymbol{y}) \propto p(\boldsymbol{\theta})f(\boldsymbol{y}|\boldsymbol{\theta}),
\end{equation*}
where $p(\boldsymbol{\theta})$ represents the prior distribution for the set of parameters $\boldsymbol{\theta}$, and $f(\boldsymbol{y}|\boldsymbol{\theta})$ stands for the likelihood of the observed data $\boldsymbol{y}$. The likelihood of the data can be rewritten as
\begin{equation*}
    f(\boldsymbol{y}|\boldsymbol{\theta})=\frac{h(\boldsymbol{y}|\boldsymbol{\theta})}{Z(\boldsymbol{\theta})},
\end{equation*}
where $Z(\boldsymbol{\theta})$ is an intractable normalizing function that makes the Metropolis-Hastings ratio impossible to calculate. By introducing an auxiliary variable $\boldsymbol{z}$ with conditional density $f(\boldsymbol{z}|\boldsymbol{y},\boldsymbol{\theta})$, we obtain an augmented target density:
\begin{equation}
    \pi(\boldsymbol{\theta}, \boldsymbol{z}| \boldsymbol{y}) \propto f(\boldsymbol{z}|\boldsymbol{y}, \boldsymbol{\theta}) p(\boldsymbol{\theta})\frac{h(\boldsymbol{y}|\boldsymbol{\theta})}{Z(\boldsymbol{\theta})}.
\end{equation}

The algorithm produces samples according to $\pi(\boldsymbol{\theta}, \boldsymbol{z}| \boldsymbol{y})$ so the $\boldsymbol{\theta}$ values correspond to samples from the marginal  distribution $\pi(\boldsymbol{\theta}| \boldsymbol{y})$, the posterior distribution of interest. \\

Consider a joint proposal for $(\boldsymbol{y}, \boldsymbol{\theta})$ that can be factorized as 
 \begin{equation}\label{eq:proposal}
         q\left(\boldsymbol{\theta}^{\prime}, \mathbf{z}^{\prime} \mid \boldsymbol{\theta}, \mathbf{z}\right)=q\left(\mathbf{z}^{\prime} \mid \boldsymbol{\theta}^{\prime}\right) q\left(\boldsymbol{\theta}^{\prime} \mid \boldsymbol{\theta}\right)=\frac{h\left(\mathbf{z}^{\prime} \mid \boldsymbol{\theta}^{\prime}\right) }{Z\left(\boldsymbol{\theta}^{\prime}\right)} q\left(\boldsymbol{\theta}^{\prime} \mid \boldsymbol{\theta}\right).  
       \end{equation}
By including the proposal in Equation (\ref{eq:proposal}) into the Metropolis-Hastings ratio $\alpha$, we cancel out the intractable normalizing function from the Metropolis-Hastings ratio and obtain the following:
\begin{equation*}
 \alpha=\min \left\{1, \frac{f\left(\mathbf{z}^{\prime} \mid \boldsymbol{\theta}^{\prime}, \mathbf{y}\right) p\left(\boldsymbol{\theta}^{\prime}\right) h\left(\mathbf{y} \mid \boldsymbol{\theta}^{\prime}\right) h(\mathbf{z} \mid \boldsymbol{\theta}) q\left(\boldsymbol{\theta} \mid \boldsymbol{\theta}^{\prime}\right)}{f(\mathbf{z} \mid \boldsymbol{\theta}, \mathbf{y}) p(\boldsymbol{\theta}) h(\mathbf{y} \mid \boldsymbol{\theta})  h\left(\mathbf{z}^{\prime} \mid \boldsymbol{\theta}^{\prime}\right)  q\left(\boldsymbol{\theta}^{\prime} \mid \boldsymbol{\theta}\right)}\right\}.
\end{equation*}
From Equation (\ref{eq:proposal}) we see that to implement the auxiliary variable MCMC method, it is necessary to sample exactly from $h\left(\mathbf{z}^{\prime} \mid \boldsymbol{\theta}^{\prime}\right) /Z\left(\boldsymbol{\theta}^{\prime}\right)$. This requires a perfect sampler, which is typically difficult or impossible to construct in general \citep[cf.][]{Park2018}. The Double Metropolis-Hastings algorithm replaces perfect sampling from the auxiliary variable $\boldsymbol{z}$ with an additional Metropolis–Hastings step. In this framework, the Metropolis–Hastings update that proposes and accepts or rejects new values of the posterior sample is called the outer sampler, whereas the Metropolis–Hastings procedure used to generate samples of the auxiliary variable $\boldsymbol{z}$ is called the inner sampler. The last state of the inner sampler is then treated like a draw from the perfect sampler in an auxiliary variable MCMC algorithm. Thus, DMH uses $m$ steps of the inner sampler to approximate perfect sampling. \\

DMH is simpler to implement and often computationally more efficient relative to other algorithms with intractable normalizing functions \cite{Park2018}. However, its computational efficiency directly depends on the efficiency of the inner sampler and the number of steps in the inner sampler $m$. The choice of the inner-sampler length $m$ is critically important, not only for computational efficiency but also because the DMH algorithm is asymptotically inexact. 
For asymptotically exact algorithms, the Markov chain’s stationary distribution is exactly equal to the target distribution. In contrast, asymptotically inexact algorithms generate Markov chains whose stationary distribution does not coincide with the target distribution; instead, they converge to an approximation of the target or, in some cases, are not known to converge to any distribution at all. In the case of DMH, the detailed balance condition for the outer sampler holds only in the limit as the inner sampler length $m \to \infty$, a condition that is impossible to achieve in practice. Consequently, it is essential to employ diagnostic tools to assess whether the resulting chains are converging to the correct target distribution. \\

\subsection{Approximate curvature diagnostic} \label{sec:ACD}

While \cite{Liang2010} and \cite{Park2018} provide  heuristics for selecting the inner-sampler length $m$, \cite{Kang2024} proposes formal diagnostics to assess whether the algorithm is approximating the target posterior distribution well. Here we consider the Approximate Curvature Diagnostic (ACD) \citep{Kang2024} which has an asymptotic $\chi^2(r)$ distribution, where $r=p_{total}(p_{total}+1) / 2$ and $p_{total}$ is the dimension of $\boldsymbol{\theta}$, if the asymptotic distribution of the sample is equal to the target posterior. Note that in the automultinomial case, $p_{total}=p \times (K-1)$. The $1-\alpha$ quantile of the $\chi^2(r)$ distribution can be used as a threshold for this diagnostic. A sample path for which the diagnostic value is below the threshold is considered to have an asymptotic distribution that is reasonably close to the target distribution. A high-level overview about this test can be found in \cite{Haran2026}, while all theoretical details can be found in \cite{Kang2024}. Because the calculation of the ACD involves approximating the gradient and Hessian of the likelihood $f(\boldsymbol{y}| \boldsymbol{\theta})$, part of the code for this calculation was implemented in \textit{JAX} \citep{jax2020}, a Python library that enables efficient automatic differentiation.

\section{Application to simulated dataset}\label{sec:simulation}

To assess the flexibility of the automultinomial model and to provide practical guidance for its application, we consider two simulated examples where  multicategory responses were simulated using alternative spatial multinomial models. The automultinomial model was fit to each simulated dataset using the DMH algorithm. To tune the algorithm and verify convergence to the right posterior distribution, we calculated the ACD until its value fell below the 99th percentile of the $\chi^2(r)$ distribution. In the first study, we generated an arrangement from a hierarchical multinomial model with Conditional Autoregressive (CAR) random effects, whereas in the second study the observed responses were generated using a mixture of Gaussian processes.

\subsection{Hierarchical model with CAR random effects}

In this simulation study, we generated a regular $30 \times 30$ grid on the unit square, for a total of $n=900$ spatial locations. Two covariates were generated independently as  $$\boldsymbol{X}_1 \sim N_{900}(2 \times \boldsymbol{1}_{900}, \boldsymbol{I}_{900}), \quad \boldsymbol{X}_2 \sim N_{900}(3 \times \boldsymbol{1}_{900},  \boldsymbol{I}_{900}).$$ 

The regression coefficient matrix used in the data-generating mechanism was

\begin{equation*}
    \boldsymbol{\beta}=\left(\begin{array}{cc}
       1  &  -0.5\\
       0.5  & 1\\
        -0.5 & -1
    \end{array}\right).
\end{equation*}
The random effects for each of the categories were generated as follows:
\begin{equation*}
    \boldsymbol{\phi}_k \sim N\left(0 \times \boldsymbol{1}_{900}, \frac{1}{0.1}\left(D-0.2A\right)^{-1}\right), 
\end{equation*}
where $A$ denotes the adjacency matrix of the regular grid, and $D$ is a diagonal matrix whose entries correspond to the number of neighbors for each grid cell. The design matrix for the model was specified as $\boldsymbol{X}=\left(\boldsymbol{1}_{900}, \boldsymbol{X_1}, \boldsymbol{X}_2\right)$. 
The categorical responses were generated using a multinomial logistic model with category $k=1$ as the reference. 
For sites $i=1,\ldots,n$ and colors $k=2, 3$, the probabilities were defined as
$$
\pi_{i k}=\frac{\exp \left(\boldsymbol{x_i}^t \boldsymbol{\beta_k} + \phi_{ik}\right)}{1+\sum_{j \neq 1} \exp \left(\boldsymbol{x_i}^t \boldsymbol{\beta_j} + \phi_{ij}\right)},
$$
while for the reference category $k=1$,
$$
\pi_{i 1}=\frac{1}{1+\sum_{j \neq 1} \exp \left(\boldsymbol{x_i}^t \boldsymbol{\beta_j} + \phi_{ij}\right)}.
$$

\begin{figure}[h]
    \centering
    \includegraphics[width=1\linewidth]{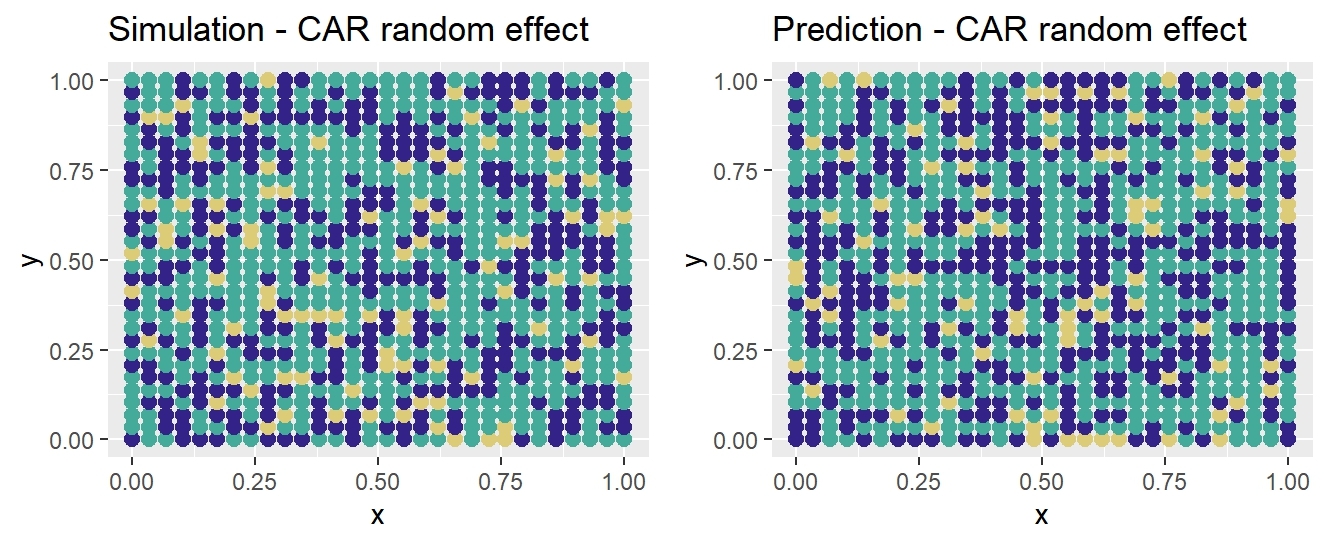}
    \caption{Simulated arrangement from hierarchical model with CAR random effects (a) and predicted arrangement under automultinomial model (b)}
    \label{fig:CAR}
\end{figure}

Figure \ref{fig:CAR} (a) shows a spatial arrangement generated under this model. To initialize the Double Metropolis–Hastings (DMH) algorithm, we used pseudolikelihood estimates obtained from the \textit{automultinomial} package. Five chains were implemented with varying inner sampler lengths ($m \in \{6,7,8,9,10\}$). Following a burn-in period of $20,000$ iterations, the Approximate Curvature Diagnostic (ACD) was progressively evaluated using chains thinned every 100 iterations. This thinning strategy was adopted to alleviate the computational cost of the ACD, which requires generating auxiliary samples for each posterior draw $\mathbf{\theta}$ to approximate the intractable normalizing function. Final diagnostics were computed after $150,000$ iterations of joint updates across all seven parameters and compared against the 99th percentile of the $\chi^2$ distribution with $28$ degrees of freedom ($\chi^2_{0.99, 28}=48.27$). The results are shown in Table \ref{tab:ACD_CAR}. We observe that the chain with the inner sampler size $m=10$ is below the diagnostic threshold, and therefore we use the corresponding output as the posterior distribution. Figure \ref{fig:CAR} (b) displays a predicted arrangement generated from the automultinomial model, parameterized by the posterior means.

\begin{table}[h]
    \centering
    \begin{tabular}{|c|c|c|c|c|c|}
    \hline
      $m$ & 6 & 7 & 8 & 9 & 10\\
      \hline
      $ACD$ & 147.20 & 76.75 & 64.55 & 77.47 & 41.37 \\
      \hline
    \end{tabular}
    \caption{ACD for different inner sample sizes in the CAR random effect study}
    \label{tab:ACD_CAR}
\end{table}

Evaluation of the predicted arrangement in Figure \ref{fig:CAR}(b) indicates that the automultinomial model effectively captures the primary structural characteristics of the original simulated dataset. Specifically, the model recovers the global prevalence of the green category and replicates the spatial configurations of all categories, including the distribution of isolated yellow sites within comparable regions of the lattice. Furthermore, the model maintains a comparable degree of spatial clustering. These results suggest that the automultinomial model is a flexible and versatile tool, capable of representing multinomial spatial patterns even when they are generated under alternative modeling specifications.

\subsection{Mixture of Gaussian Processes} \label{sec:SimGP}
\begin{figure}
    \centering
    \includegraphics[width=0.4\linewidth]{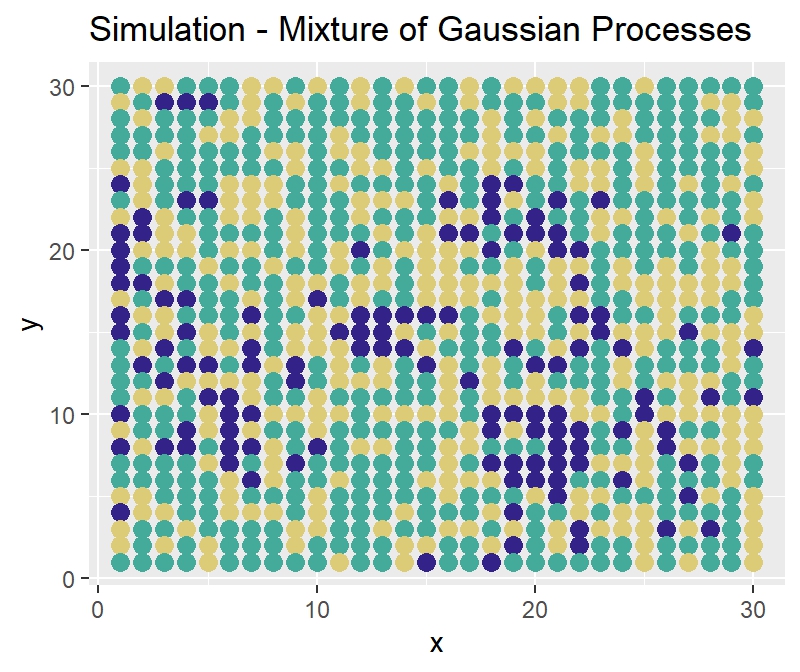}
    \caption{Simulated arrangement from mixture of Gaussian Processes}
    \label{fig:OriginalGP}
\end{figure}

For this analysis, we simulated a multicategorical arrangement resulting from a mixture of Gaussian processes with the goal of studying the behavior of the automultinomial model in smoother arrangements with higher levels of spatial clustering. To do so, we generated three Gaussian Processes as follows: 

\begin{enumerate}
    \item We simulated two covariates as $X_1 \sim N(0, 2)$ and $X_2 \sim N(1, 3)$. We did not include an intercept in this model.
    \item We calculated the Euclidean distance between the centroids of the cells. The distance between the centroids of cell $i$ and cell $j$ will be denoted by $d_{ij}$.
    \item For $K=3$ classes, we simulated three different Gaussian processes $\mathbf{Z}_k$, $k=1,2,3$ using the $\gamma-\text{exponential}$ kernel with length parameter $l$ to model spatial correlation. That is $$\mathbf{Z}_k \sim GP(\boldsymbol{\mu}_k, \Sigma),$$ where
    $$\boldsymbol{\mu}_k=\boldsymbol{X}\boldsymbol{\beta}_{k},$$
    $$\boldsymbol{\beta}=\left(\boldsymbol{\beta}_1, \boldsymbol{\beta}_2, \boldsymbol{\beta}_3\right)=\left(        \begin{tabular}{ccc}
           0 &  -0.5 & 2\\
           0  & 1 & -1\\
        \end{tabular}\right),$$

    and, 
    $$\Sigma^{(i,j)}=\exp\Bigl\{-\left(d_{ij}/l\right)^\nu\Bigl\},$$
    $l=3$ and $\nu=2$.
    \item We assigned the value $k=\max_{i}\{Z_i, i=1,\ldots, 4\}$ to each cell.  
\end{enumerate}
The resulting $30 \times 30$ arrangement is presented in Figure \ref{fig:OriginalGP}.\\

In this simulation study, we ran chains of the DMH algorithm with inner sample sizes $m=3, 4, \ldots, 15$ for different outer lengths. To improve mixing of the Double Metropolis–Hastings algorithm, we adopted a block-updating strategy. Specifically, one block was assigned to the parameters associated with class 2, a second block to the parameters associated with class 3, and a third block to the spatial correlation parameter. The covariance matrices used for the multivariate normal proposal distributions in each block were obtained by scaling the covariance matrices derived from the pseudolikelihood estimates, with the aim of achieving acceptance rates between $20$ to $30\%$ for blocks 1 and 2, and between $30$ to $40\%$ for block 3. The calculated ACD values obtained in each case were compared to the 99th percentile of a $\chi^2$ distribution with 15 degrees of freedom ($\chi^2_{0.99, 15}=30.58$). Some of the calculated ACD values are presented in Table \ref{tab:ACD_GP}.

\begin{table}[h]
    \centering
    \begin{tabular}{|c|c|c|c|c|c|c|}
    \hline
      $m$ & 3 & 7 & 9 & 11 & 13 & 15\\
      \hline
      $ACD$ & 74.76 & 63.83 & 75.14 & 112.02 & 63.47 & 84.06\\
      \hline
    \end{tabular}
    \caption{ACD for different inner sample sizes in the Gaussian processes study}
    \label{tab:ACD_GP}
\end{table}

We observe that none of the inner sample sizes considered achieved ACD values below the prescribed threshold. It is important to emphasize that the ACD is a diagnostic designed to evaluate convergence of the algorithm to a specified target distribution, in this case, the automultinomial model, rather than a measure of goodness of fit or an indication of model misspecification. To further investigate the behavior of the algorithm, we report the posterior means and corresponding  95\% probability intervals associated with the runs used to compute the ACD values in Table~\ref{tab:ACD_GP}. These summaries are presented in Table~\ref{tab:resultsGP}. \\ 

Although the ACD criterion is not satisfied, posterior summaries remain stable across inner sample sizes, suggesting that the DMH sampler yields consistent posterior approximations in practice. From this perspective, the algorithm may be viewed as asymptotically inexact in this setting, yet empirically robust with respect to the inferential quantities of interest. Understanding the persistent ACD values requires further targeted investigation; plausible explanations include limited efficiency of the auxiliary chain in exploring high-probability regions or the possibility that the ACD is overly conservative for strongly dependent spatial models.\\

\begin{table}[h]
\centering
\begin{tabular}{lrrrr}
\hline
Parameter & Inner & Post. & \multicolumn{2}{c}{95\% CI} \\
          & sample & mean & Lower & Upper \\
\hline
\multirow{6}{*}{$\beta_{21}$}
 & 3  & -0.48 & -0.68 & -0.28 \\
 & 7  & -0.47 & -0.67 & -0.27 \\
 & 9  & -0.47 & -0.67 & -0.25 \\
 & 11 & -0.46 & -0.73 & -0.24 \\
 & 13 & -0.47 & -0.66 & -0.28 \\
 & 15 & -0.48 & -0.70 & -0.28 \\
\hline
\multirow{6}{*}{$\beta_{22}$}
 & 3  & 1.17 & 0.97 & 1.38 \\
 & 7  & 1.19 & 0.99 & 1.40 \\
 & 9  & 1.17 & 0.97 & 1.37 \\
 & 11 & 1.18 & 1.02 & 1.38 \\
 & 13 & 1.17 & 0.98 & 1.35 \\
 & 15 & 1.17 & 1.00 & 1.36 \\
\hline
\multirow{6}{*}{$\beta_{31}$}
 & 3  & 2.69 & 2.29 & 3.09 \\
 & 7  & 2.70 & 2.28 & 3.18 \\
 & 9  & 2.72 & 2.28 & 3.26 \\
 & 11 & 2.70 & 2.28 & 3.21 \\
 & 13 & 2.69 & 2.26 & 3.17 \\
 & 15 & 2.67 & 2.18 & 3.15 \\
\hline
\multirow{6}{*}{$\beta_{32}$}
 & 3  & -1.35 & -1.65 & -1.03 \\
 & 7  & -1.35 & -1.67 & -1.04 \\
 & 9  & -1.36 & -1.70 & -1.07 \\
 & 11 & -1.35 & -1.69 & -1.04 \\
 & 13 & -1.33 & -1.63 & -1.04 \\
 & 15 & -1.33 & -1.64 & -1.04 \\
\hline
\multirow{6}{*}{$\gamma$}
 & 3  & 0.40 & 0.17 & 0.62 \\
 & 7  & 0.41 & 0.18 & 0.60 \\
 & 9  & 0.40 & 0.20 & 0.60 \\
 & 11 & 0.41 & 0.22 & 0.61 \\
 & 13 & 0.40 & 0.16 & 0.63 \\
 & 15 & 0.39 & 0.19 & 0.59 \\
\hline
\end{tabular}
    \caption{Posterior means and $95\%$ probability intervals obtained from DMH using different inner sample sizes}
    \label{tab:resultsGP}
\end{table}

\begin{figure}[h]
    \centering
    \includegraphics[width=1\linewidth]{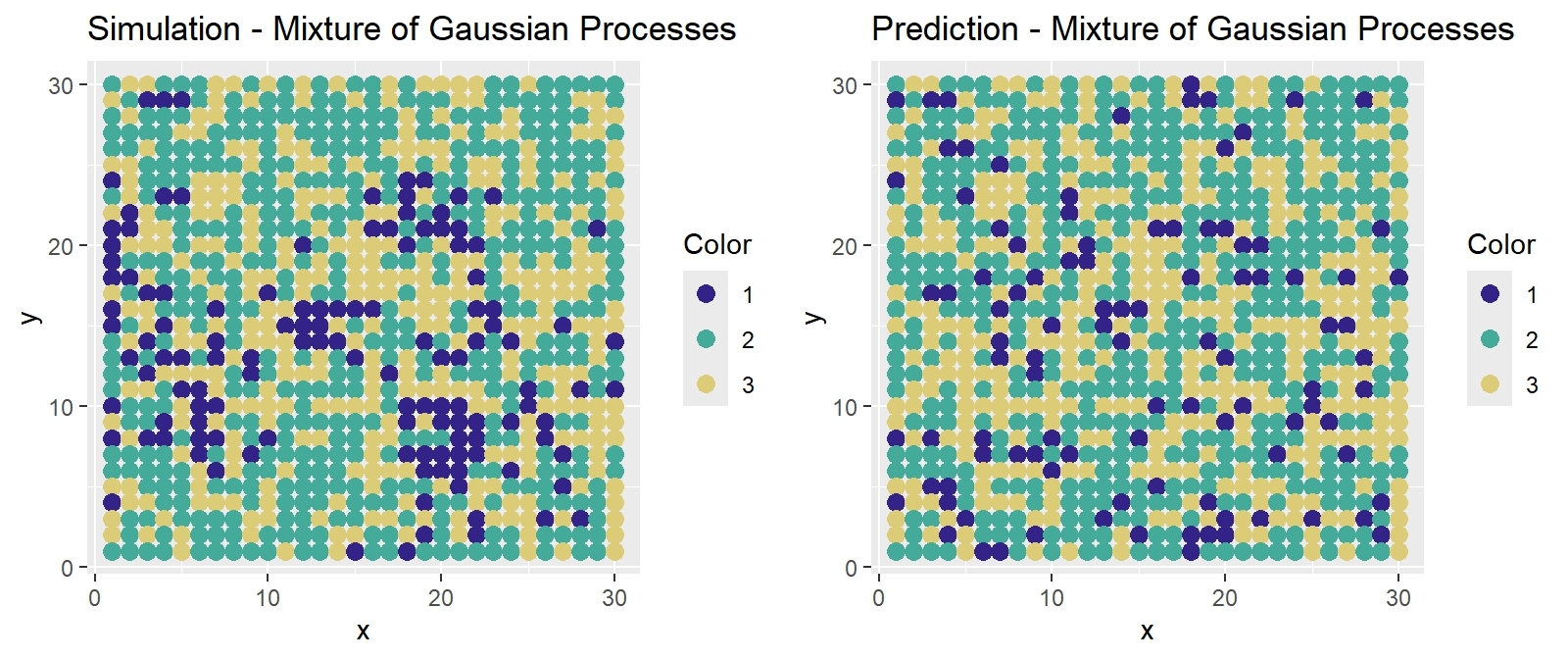}
    \caption{Simulated arrangement from mixture of Gaussian Processes (a) and predicted arrangement under automultinomial model (b)}
    \label{fig:GPResult}
\end{figure}

Finally, Figure~\ref{fig:GPResult} presents the original data alongside a prediction obtained using the posterior means from the DMH algorithm with inner sample size 
$m=13$. The model reproduces the main spatial patterns for colors 2 and 3 reasonably well, although it shows more difficulty capturing the clustering structure associated with color 1. One possible explanation is the comparatively smaller number of observations for this category relative to the others, which may limit the model’s ability to learn its spatial dependence structure. Overall, however, we find that the automultinomial model provides a reasonable representation of data generated from a flexible non-automultinomial mechanism, suggesting that it can serve as a flexible and versatile tool for modeling multicategory areal data.

\section{Land cover data application} \label{sec:dataanalysis}

In this section, we apply the automultinomial model to satellite-derived land cover observations from a region of Southeast Asia bounded by $17^{\circ}$ to $19^{\circ}$ N and $98.5^{\circ}$ to $105^{\circ}$ E for the year 2005. Deforestation is a major concern in Southeast Asia, with a significant percentage of the original forests lost to other land uses  \citep{Berrett2010}. Understanding the associations between land cover types and key predictors such as elevation and proximity to major roads, coastlines, and urban centers can provide valuable insights for researchers examining the socioeconomic and demographic drivers of land-use change. Given the categorical data and the presence of spatial dependence, the automultinomial model is appropriate for characterizing covariate effects while accounting for spatial correlation. Additional details regarding the dataset and its source are provided in \citet{Berrett2010}. \\

\begin{figure}
    \centering
    \includegraphics[width=0.45\linewidth]{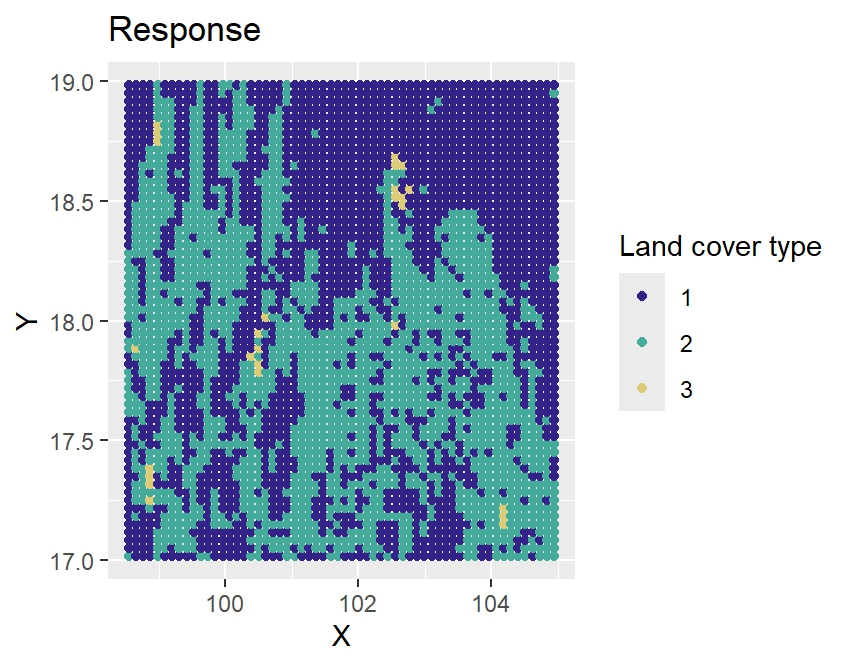}
    \caption{Land cover in the region of study}
    \label{fig:response}
\end{figure}

\begin{figure}
    \centering
    \includegraphics[width=0.7\linewidth]{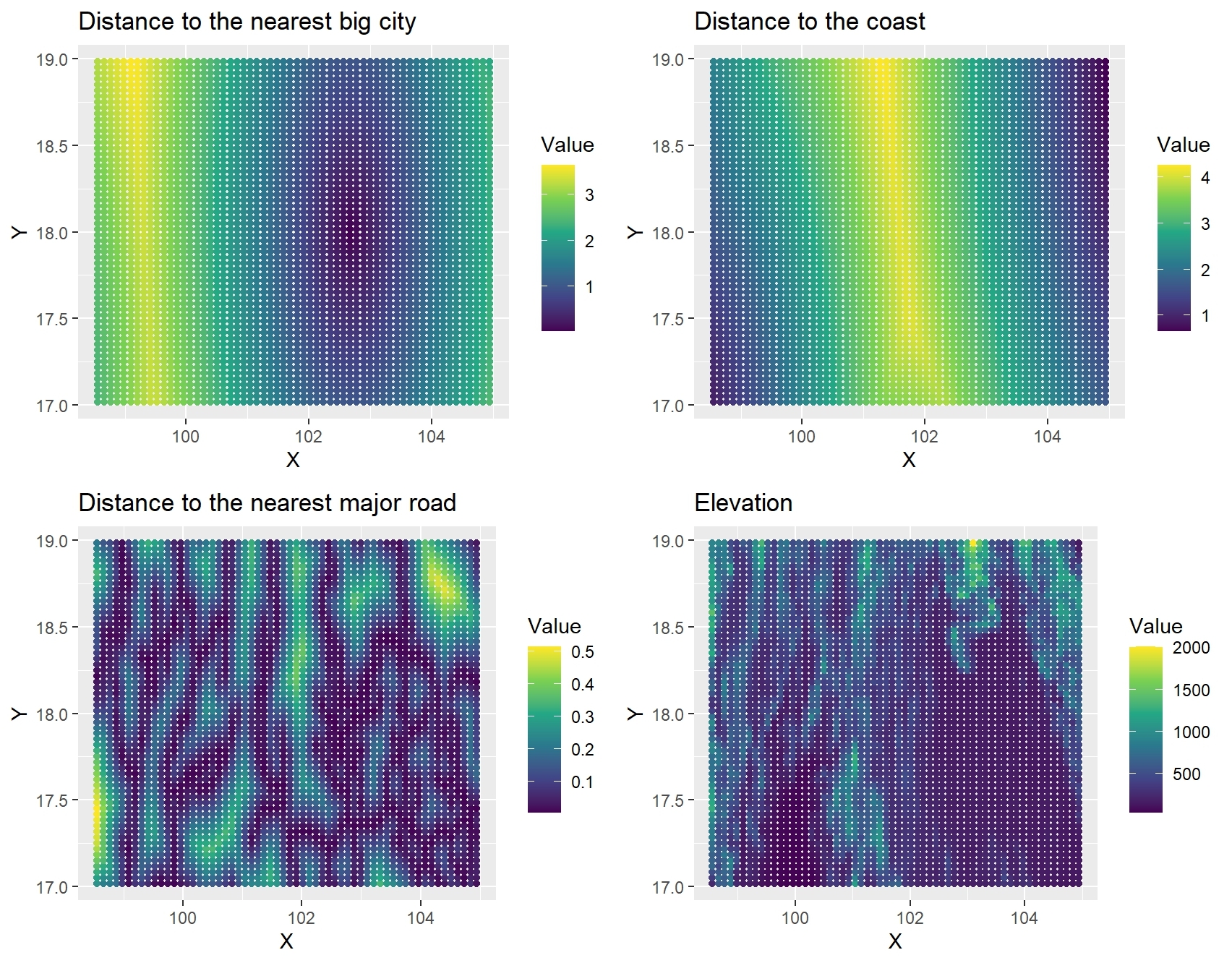}
    \caption{Covariates in the region of study}
    \label{fig:covariates}
\end{figure}

To prepare the data for analysis with the automultinomial model, we overlaid a regular $60 \times 60$ grid on the study region, resulting in a total of $3600$ areal units. The International Human Dimensions Programme (IHDP) land cover classes in the original dataset were aggregated into three categories: forest, non-forest, and other, where the latter includes wetland, urban, barren, and water classes.
Because each grid cell contains multiple observation points, the data were summarized at the cell level. For the land cover response, we computed the proportion of points belonging to each of the three categories within a cell and assigned the category with the highest proportion. For each of the four predictors, the cell-level covariate value was taken to be the mean of the observations within that cell. Figure~\ref{fig:response} presents the resulting land cover arrangement, while Figure~\ref{fig:covariates} displays the covariates over the study region. \\

In this case, the model consists of three classes, with class 1 as the reference category. Given four covariates and an intercept, this results in a total of 11 regression coefficients to be estimated across the remaining classes including the spatial correlation parameter $\gamma$. To improve mixing of the Double Metropolis–Hastings algorithm, we adopted the same block-updating strategy described in Section \ref{sec:SimGP}. \\

With an inner sampler length of $m=8$, we ran the chain for a burn-in period of $20000$ iterations. The ACD was then evaluated using chains until the diagnostic value fell below the prescribed threshold; the chain was thinned every $300$ iterations. In this setting, the covariance matrix appearing in the ACD calculation was not full rank. This behavior is expected and arises from linear dependencies induced by the identifiability constraints inherent to the automultinomial model, such as the use of a reference category and the resulting redundancy among parameters. Although this singularity does not allow direct inversion of the covariance matrix, it does not prevent computation of the ACD statistic. We therefore employed a generalized inverse in place of the standard matrix inverse. Consequently, the diagnostic threshold was taken to be the 99th percentile of a  $\chi^2$ distribution with 26 degrees of freedom, corresponding to the rank of the covariance matrix used in the calculation. After $53,000$ iterations, the ACD yielded a value of $27.31$, which is below the threshold of $45.64$. Table~\ref{tab:results} presents the posterior means for all model parameters together with their approximate $95\%$ posterior probability intervals.\\

\begin{table}
    \centering
    \begin{tabular}{|c|c|c|c|}
    \hline
      \textbf{Parameter}   & \textbf{Posterior mean} & \textbf{Lower bound} & \textbf{Upper bound}\\
      \hline
      Intercept - Non-forest   & 0.003 & -0.239 & 0.242 \\
         \hline
       Elevation - Non-forest  & -0.002 & -0.003 & -0.001\\
         \hline
        Distance to coast - Non-forest &  0.248 & 0.149 & 0.339\\
         \hline
        Distance to city - Non-forest & 0.283 & 0.203 & 0.379\\
         \hline
       Distance to roads - Non-forest  & -3.357 & -4.268 & -2.377\\
         \hline
       Intercept - Other   & -2.009 & -3.649 & 0.030\\
         \hline
       Elevation - Other  & -0.005 & -0.007 & -0.002\\
         \hline
        Distance to coast - Other & -0.088 & -0.627 & 0.420\\
         \hline
        Distance to city - Other & 0.167 & -0.282 & 0.588\\
         \hline
       Distance to roads - Other  & 1.046 & -3.441 & 5.022\\
         \hline
        Spatial correlation & 0.875 & 0.816 & 0.939 \\
        \hline
    \end{tabular}
    \caption{Posterior means and $95\%$ posterior probability intervals for automultinomial parameters}
    \label{tab:results}
\end{table}

\begin{figure}
    \centering
    \includegraphics[width=1\linewidth]{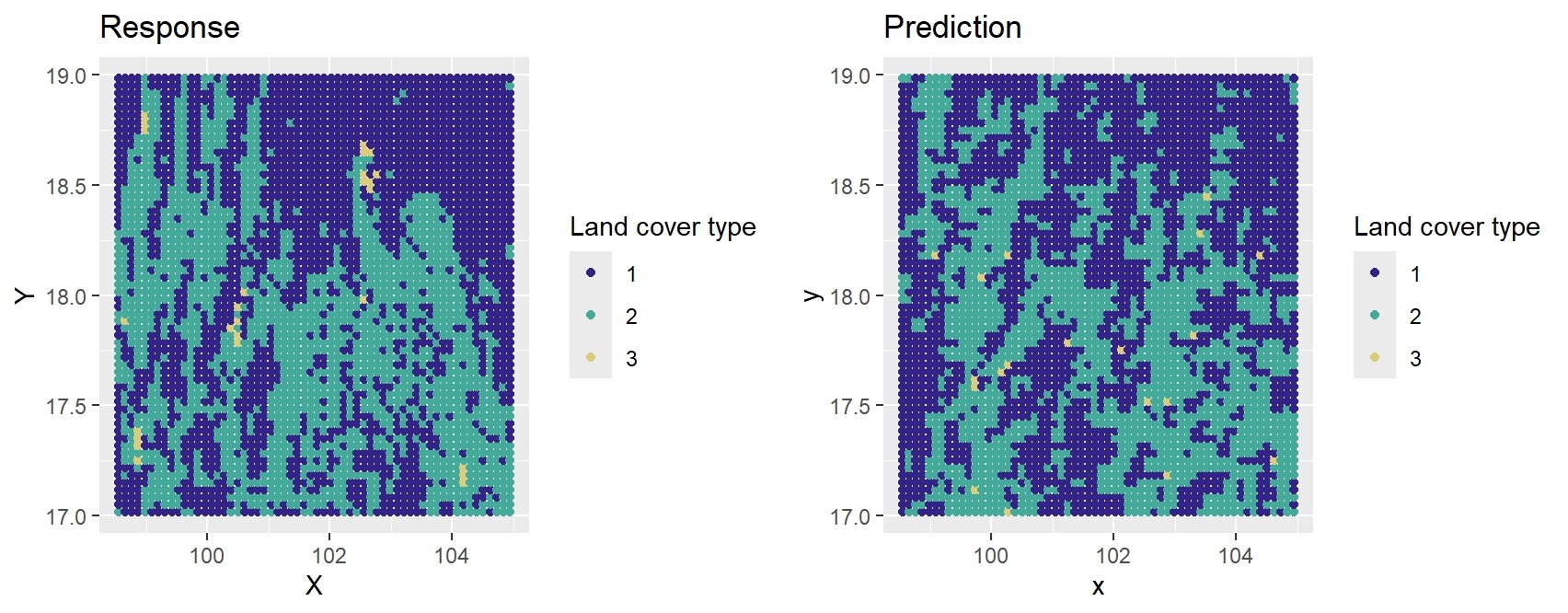}
    
    \caption{Original and prediction of land cover in the region of study}
    \label{fig:prediction}
\end{figure}

Finally, Figure~\ref{fig:prediction} presents a predicted arrangement generated from the automultinomial model, parameterized by the posterior means. Overall, the model successfully recovers the primary structural features of the observed landscape, including broad spatial trends and the degree of spatial autocorrelation. However, the model encounters greater difficulty in accurately placing cells belonging to class 3. This result is expected, given the relative sparsity of observations for this category, which limits the information available for parameter estimation. Additionally, the model over-predicts the presence of non-forest cells in the northeast quadrant of the study area. These discrepancies suggest that while the automultinomial framework captures global spatial dependencies well, it may be sensitive to localized variations or class imbalances within the dataset.

\section{Discussion} \label{sec:discussion}

In this work, we evaluated the automultinomial model as a potentially useful framework for modeling multicategory responses on a lattice. Based on both simulation studies and an application to real data, we found that the automultinomial model is flexible across a wide range of spatial correlation levels. Moreover, the use of a single parameter to characterize spatial dependence simplifies model specification. However, certain limitations were observed in the localized accuracy of the model. In the real data application, the over-prediction of non-forest cells in the northeast quadrant and the difficulty in localizing the sparse ``class 3'' category suggest that the model may struggle with significant class imbalances. In the context of land-change modeling, minority classes often represent critical transitions. In this instance, Class 3 includes urban and barren land uses. The scarcity of these observations presents a substantial practical challenge for parameter estimation. \\

We found the DMH algorithm to be a practical and relatively straightforward approach for posterior inference in the presence of the intractable normalizing function. The ACD proved to be a valuable tool for assessing the quality of posterior samples and for determining whether the Markov chain had reached the target posterior distribution. Although DMH can be computationally demanding in high-dimensional settings, this cost is closely tied to the efficiency of the sampler. In such cases, we recommend the use of block-updating strategies to improve mixing and overall performance. 
An advantage of the ACD is that its computation can be embarrassingly parallel. While the approximation of gradients and Hessians may be computationally intensive, we found the JAX library in Python to be highly efficient for these tasks. Consequently, the overall computational burden of the diagnostic is primarily driven by the cost of generating the auxiliary samples required for its evaluation, which again depends on the efficiency of the underlying sampler. As a limitation, the current implementation relies on careful tuning of the inner sampler length and proposal distributions, which may become increasingly challenging as the dimension of the parameter space grows. \\

Future research may explore a centered parameterization for the automultinomial model to enhance the interpretability of its parameters. \citet{caragea2009} presented this parameterization for the autologistic model aiming to decouple the large-scale (systematic) and small-scale (spatial) components of the model. In standard Markov Random Field specifications, the intuitive large-scale component does not directly correspond to marginal expectations; conversely, the small-scale component does more than simply introduce local modification but it often fundamentally shifts the marginal means. By implementing a centered framework, researchers could more effectively isolate the effects of covariates from the underlying spatial dependence. Later work by \citet{Kaiser2012} generalized the centered parameterization for a broad class of Markov Random Field models (auto-models), the explicit derivation and implementation for the multinomial case remains to be addressed. Specifically, extending these centering techniques to handle the multi-category dependencies and the associated constraints of the automultinomial framework brings unique challenges that surpass those found in Gaussian, binary, or count-based auto-models. Addressing this gap would facilitate more precise estimation of covariate effects. On the other hand, evaluating the goodness of fit of the automultinomial model is also an open problem, as previous works have focused on evaluating models with continuous responses \citep{Kaiser2012b} and binary responses \citep{Biswas2024}. \\

\section{Acknowledgments}

We thank Dr.\ Candace Berrett for her generosity in providing the dataset used in the application section of this work.

\section{Data availability statement}

The data and reproducible code that support the findings of this study are available from the corresponding author upon reasonable request.

 \section{Funding statement}

The authors declare that no funds, grants, or other support were received during the preparation of this manuscript.

\section{Conflict of interest statement}

The authors report there are no competing interests to declare.

\bibliography{refs}

@inproceedings{Geyer1991,
  title        = {Markov {C}hain {M}onte {C}arlo {M}aximum {L}ikelihood},
  author       = {Charles J. Geyer},
  year         = 1991,
  booktitle    = {Computing Science and Statistics, Proceedings of the 23rd Symposium on the Interface},
  pages        = {156--163}
}

@article{Potts1952,
author = {R. B. Potts},
title = {Some Generalized Order-Disorder Transformations},
journal = {Mathematical Proceedings of the Cambridge Philosophical Society},
volume = {48},
number = {1},
pages = {106--109},
year = {1952}
}

@book{Cressie,
    title = {Statistics for Spatial Data},
    author = {Noel A. C. Cressie},
    isbn = {9780471002550},
    series = {Wiley Series in Probability and Statistics},
    year = {1993},
    publisher = {Wiley}
}

@book{Zuur2007,
    title = {Analyzing Ecological Data},
    author = {Alain F. Zuur and Elena N. Ieno and Graham M. Smith},
    isbn = {9780387459677},
    year = {2007},
    series = { Statistics for Biology and Health},
    publisher = {CRC Press}
}

@article{Hughes2013,
  title={Dimension Reduction and Alleviation of Confounding for Spatial Generalized Linear Mixed Models},
  author={John Hughes and Murali Haran},
  journal={Journal of the Royal Statistical Society Series B: Statistical Methodology},
    volume = {75},
  issue = {1},
  pages = {139–159},
 year = {2013}
}

@article{Haran2025,
  title={Latent {G}aussian Models and Computation for Large Spatial Data},
  author={Murali Haran and John Hughes and Ben Seiyon Lee},
  journal={To appear in Handbook of MCMC, 2nd Edition},
 year = {2025+}
}

@article{Hughes2011,
  title={Autologistic models for binary data on a lattice},
  author={John Hughes and Murali Haran and Petruţa C. Caragea},
  journal={Environmetrics},
    volume = {22},
  issue = {7},
  pages = {857-871},
 year = {2011}
}

@article{rue2009approximate,
  title={Approximate Bayesian inference for latent Gaussian models by using integrated nested Laplace approximations},
  author={Rue, H{\aa}vard and Martino, Sara and Chopin, Nicolas},
  journal={Journal of the Royal Statistical Society Series B: Statistical Methodology},
  volume={71},
  number={2},
  pages={319--392},
  year={2009},
  publisher={Oxford University Press}
}

@article{Berrett2012,
  title = {Data augmentation strategies for the {B}ayesian spatial probit regression model},
  author = {Candace Berrett and Catherine Calder},
  journal = {Computational Statistics and Data Analysis},
    volume = "56",
    pages = "478--490",
  year = {2012}
}

@phdthesis{Berrett2010,
  title        = {Bayesian Probit Regression Models for Spatially-Dependent Categorical Data},
  author       = {Candace Berrett},
  year         = {2010},
  note         = {Available at \url{https://etd.ohiolink.edu/acprod/odb_etd/etd/r/1501/10?clear=10&p10_accession_num=osu1285076512}},
  school       = {The Ohio State University},
  type         = {Ph{D} thesis}
}

@article{Imai2005,
  title = {A {B}ayesian analysis of the multinomial probit model using marginal data augmentation},
  author = {Kosuke Imai and David A. van Dyk},
  journal = {Journal of Econometrics},
    number = "124",
    pages = "311--334",
  year = {2005}
}

@article{Jin2013,
  title = {Spatial multinomial regression models for nominal categorical data: a study of land cover in {Northern Wisconsin, USA}},
  author = {Chongyang Jin and Jun Zhub and Michelle M. Steen-Adams and Stephan R. Sain and Ronald E. Gangnon},
  journal = {Environmetrics},
    number={24},
    pages = "98--108",
  year = {2013}
}

@article{Besag1974,
  title = {Spatial Interaction and the Statistical Analysis of Lattice Systems},
  author = {Julian Besag},
  journal = {Journal of the Royal Statistical Society. Series B (Methodological)},
volume={36},
    number = "2",
    pages = "192--236",
  year = {1974}
}

@article{Besag1975,
  title = {Statistical Analysis of Non-Lattice Data},
  author = {Julian Besag},
  journal = {The Statistician},
volume={24},
    number = "3",
    pages = "179--195",
  year = {1975}
}

@article{besag1991bayesian,
  title={Bayesian image restoration, with two applications in spatial statistics},
  author={Besag, Julian and York, Jeremy and Molli{\'e}, Annie},
  journal={Annals of the institute of statistical mathematics},
  volume={43},
  pages={1--20},
  year={1991},
  publisher={Springer}
}

@article{diggle1998model,
  title={Model-based geostatistics},
  author={Diggle, Peter J and Tawn, Jonathan A and Moyeed, Rana A},
  journal={Journal of the Royal Statistical Society Series C: Applied Statistics},
  volume={47},
  number={3},
  pages={299--350},
  year={1998},
  publisher={Oxford University Press}
}

@article{DuenasH2025,
   title={Modeling discrete lattice data using the Potts and tapered Potts models},
   author={Duenas-Herrera, Maria Paula and Berg, Stephen and  Haran, Murali},
   journal={arXiv preprint arXiv:2509.21478},
   year={2025}
}

@article{Park2018,
  title = {Bayesian Inference in the Presence of Intractable Normalizing Functions},
  author = {Jaewoo Park and Murali Haran},
  journal = {Journal of the American Statistical Association},
volume={13},
    number={523},
    pages = "1372--1390",
  year = {2018}
}

@article{Liang2010,
  title = {A double Metropolis–Hastings sampler for spatial
models with intractable normalizing constants},
  author = {Faming Liang},
  journal = {Journal of Statistical Computation and
Simulation},
volume={80},
    number={9},
    pages = "1007--1022",
  year = {2010}
}

@article{Kang2024,
  title = {Measuring Sample Quality in Algorithms for Intractable Normalizing
Function Problems},
  author = {Bokgyeong Kang and John Hughes and Murali Haran},
  journal = {Journal of Machine Learning Research},
    number={25},
    pages = "1--32",
  year = {2024}
}

@misc{Berg2019_2,
  author = {Stephen Berg},
  title = {Automultinomial},
  year = {2019},
  publisher = {GitHub},
  journal = {GitHub repository},
  url = {https://github.com/stephenberg/automultinomial/blob/master/vignettes/vignette.pdf},
  note = {[Online; accessed December 15, 2025]}
}

@article{Moller2006,
  title = {An Efficient Markov Chain Monte Carlo Method for Distributions with Intractable Normalising Constants},
  author = {J. Møller and A. N. Pettitt and R. Reeves and K. K. Berthelsen},
  journal = {Biometrika},
  volume = {93},
    number={2},
    pages = "451--458",
  year = {2006}
}

@article{haran2011gaussian,
  title={Gaussian random field models for spatial data},
  author={Haran, Murali},
  journal={Handbook of Markov Chain Monte Carlo},
  pages={449--478},
  year={2011},
  publisher={CRC Press Boca Raton, FL, USA}
}

@article{Haran2026,
  title={Algorithms for Models with Intractable Normalizing Functions},
  author={Murali Haran and Bokgyeong Kang and Jaewoo Park},
  journal={To appear in Handbook of MCMC, 2nd Edition},
 year = {2026+}
}

@misc{jax2020,
  title        = {JAX: High-Performance Machine Learning Research},
  author       = {{JAX Developers}},
  year         = {2020},
  howpublished = {\url{https://github.com/google/jax}}
}

@article{caragea2009,
  title        = {Autologistic Models With Interpretable Parameters},
  author       = {Caragea, Petru\c{t}a C. and Kaiser, Mark S.},
  journal      = {Journal of Agricultural, Biological and Environmental Statistics},
  volume       = {14},
  number       = {3},
  pages        = {281--300},
  year         = {2009},
  doi          = {10.1198/jabes.2009.07032}
}

@article{Kaiser2012,
  title        = {Centered parameterizations and dependece limitations in Markov random field models},
  author       = {Kaiser, Mark S. and Caragea, Petru\c{t}a C. and Furukawa, Kyoji},
  journal      = {Journal of Statistical Planning and Inference},
  volume       = {142},
  number       = {7},
  pages        = {1855--1863},
  year         = {2012},
  doi          = {10.1016/j.jspi.2012.02.030}
}

@article{Kaiser2012b,
  title        = {Goodness of fit tests for a class of markov random field models},
  author       = {Kaiser, Mark S. and Lahiri, Soumendra and Nordman, Daniel J.},
  journal      = {The Annals of Statistics},
  volume       = {40},
  number       = {1},
  pages        = {104--130},
  year         = {2012},
  doi          = {10.1214/11-AOS948}
}

@article{Biswas2024,
  title        = {A formal goodness- of- fit test for spatial binary Markov random field models},
  author       = {Biswas, Eva and Kaplan Andee and Kaiser, Mark S. and Nordman, Daniel J.},
  journal      = {Biometrics},
  volume       = {80},
  number       = {4},
  pages        = {1--11},
  year         = {2024},
  doi          = {10.1093/biomtc/ujae119}
}

@article{Hodges2012,
  title        = {Adding Spatially-Correlated Errors Can Mess Up the Fixed Effect You Love},
  author       = {Hodges, James S. and Reich, Brian J.},
  journal      = {The American Statistician},
  volume       = {64},
  number       = {4},
  pages        = {325--334},
  year         = {2012},
  doi          = {10.1198/tast.2010.10052}
}

@article{Hughes2017,
  title        = {Spatial Regression and the Bayesian Filter},
  author       = {John Hughes},
    journal={arXiv preprint arXiv:1706.04651},
   year={2017}
}
\end{document}